\documentclass{aastex62}

\graphicspath{{./}{figures/}}

\received{January 1, 2018}
\revised{January 7, 2018}
\accepted{\today}
\submitjournal{ApJ}

\shorttitle{Sample article}
\shortauthors{Schwarz et al.}

\begin{document}

\title{A new solution for the mass growth of Black Holes consistent with thermodynamics
}

\correspondingauthor{M. F. La Rotta Wilches}
\email{mf.larottawilches@ugto.mx}

\author[0000-0003-3446-4025]{Mar\'{\i}a F. La Rotta Wilches}
\affil{Departamento de Astronom\'{\i}a, Universidad de Guanajuato\\ Callejón de Jalisco S/N; Valenciana; C.P. 36240; \\ Guanajuato, Gto. México. }

\author[0000-0001-6927-522X]{Roger Coziol}
\affil{Departamento de Astronom\'{\i}a, Universidad de Guanajuato\\ Callejón de Jalisco S/N; Valenciana; C.P. 36240; \\ Guanajuato, Gto. México. }



\begin{abstract}
The James Webb Space Telescope (JWST) has recently revealed evidence of supermassive black holes (SMBHs) forming in the cores of high redshift galaxies that grew in mass extremely rapidly during the first billion years of the universe. In this study we present a new solution where the growth mass factor of BHs, $\gamma(t) \propto \dot M_{BH}/M_{BH}$, varies with time, following the changes of radiative efficiency, $\epsilon$, and Eddington ratio, $\lambda_{Edd}$, caused by the variation in accretion flow. By optimizing the accretion rate, $\dot M_{BH}$, we obtained a solution that is universal, that is, which applies to any BH at any redshift. This suggests that the mass of the BH is a thermodynamics state function, in good agreement with the no hair theorem and the four laws of mechanics of BHs.  
\end{abstract}



\section{Introduction} 
\label{sec:intro}

Before the era of the James Web Space Telescope (JWST) the model for the formation of super massive black holes (SMBHs) in active galactic nuclei (AGN) was thought to be straightforward. Following the hierarchical structure formation scenario, massive galaxies form by the mergers of gas-rich protogalaxies, where clusters of population~III (PopIII) stars rapidly evolve and merge into SMBH seeds, with typical masses $10^3-10^5$ M$_\odot$ \citep{Klessen2023}, which then grow in mass trough accretion of matter, emitting copious amount of radiation through their accretion disks. Adding to this view the observational facts that, 1- the population of quasi stellar objects (QSOs) or QUASARs, their radio-loud version \citep[which are less common][]{Coziol2017}, with BH masses $M_{BH} \sim 10^{8-9}$~M$_\odot$ become less conspicuous and luminous at low-$z$, and 2- the masses of SMBHs in the bulges of galaxies at low $z$ are correlated with the velocity dispersions of the stars, the simplest conclusion was that the formation of all galaxies must pass by an AGN phase \citep{Sanders1988,Bahcall1997,CutivaAlvarez2023}.

Then came the JWST, although not contradicting totally this rational but seriously challenging our understanding of how the whole picture can make sense physically \citep{Corredoira2024,Conselice2025,Jockel2026}. More specifically, JWST discovered a population of plausible high-$z$ galaxies where star formation is extremely high, dominating over the AGN activity, suggesting that galaxies might have formed before their SMBHs \citep{Wang2024,Terrazas2025}. It also uncovered a rich population of unresolved and very red sources, little red dots (LRDs), appearing only at $z \gtrsim4$, suggesting SMBH are growing in mass hidden behind a veil of dust  \cite{Matthee2024,Li2025,Akins2025,Tanaka2025}, and, when SMBHs are clearly observed, their masses seem unusually high compared to their host galaxies  \citep{Hegde2024,Tripodi2025,Jeon2025}, implying their growth was extremely fast, taking just a few hundred million years to reach $10^7-10^9$ M$_\odot$. This either suggest BH seeds were supermassive, $10^6-10^7$ M$_\odot$ \citep{Sanati2025}, or less massive, $\lesssim 10^4$ M$\odot$, but followed by a phase of super-Eddington accretion \cite{Aggarwal2025,Inayoshi2025}. Although the task of explaining these extraordinary observations seems daunting, there are some suggestions that the solution, in fact, could be easier than thought \citep[e.g.,][]{Jahnke2025}. To realize that, however, one must study in detail the physics of accretion onto a SMBH \citep[e.g.][]{Aggarwal2025}. 

\section{Reconstructing the accretion process in details}

Following \citet[][]{King2006} a fast growth of SMBH is easily obtained assuming the accretion disk radiative efficiency--that is, how efficiently matter is transformed into light--is significantly reduced. Since the luminosity of the observed AGN is given by:
\begin{equation}\label{eqn:LAGN}
L_{AGN} = \epsilon \dot{M}_{acc}c^2
\end{equation}
where $c$ is the light velocity, $\dot{M}_{acc}$, the accretion rate of matter onto the disk and $\epsilon$ the radiative efficiency, the BH mass itself must increase as:
\begin{equation}\label{eqn:Mdot}
\dot{M}_{BH} = (1-\epsilon) \dot{M}_{acc}
\end{equation}
This implies that the lower is $\epsilon$, the higher the mass accreting onto the BH. 

The accretion process, on the other hand, is limited by two physical parameters: 1- $\epsilon$, which cannot be unity, because that would imply all the mass falling onto the BH is transformed into light without producing entropy, violating the second law of thermodynamics \citep{Bekenstein1972}, and 2- the Eddington luminosity, $L_{Edd}$, which has the remarkable property that it depends only on two quantities, the mass of the BH and the electron scattering opacity, $\kappa$: 
\begin{equation}\label{eqn:LEdd}
L_{Edd} = \frac{4\pi G}{\kappa c}M_{BH}c^2=\frac{M_{BH} c^2}{t_{Edd}}
\end{equation}
where $t_{Edd}= 4.5 \times 10^8$ years is the Eddington time. Consequently, the Eddington ratio itself, that is, the ratio of the bolometric luminosity to the Eddington luminosity, $\lambda_{Edd} = L_{bol}/L_{Edd}$, cannot be higher than unity (unless, contrary to what Eddington assumed, the emission is non-isotropic).

Combining Eqn~\ref{eqn:LAGN}, Eqn.~\ref{eqn:Mdot} and Eqn~\ref{eqn:LEdd}
one can eliminate $\dot{M}_{acc}$, obtaining a simple differential equation with separable variables:
\begin{equation}\label{eqn:diffEqn}
\frac{dM_{BH}}{M_{BH}}= \lambda_{Edd}\frac{(1-\epsilon)}{\epsilon}\frac{dt}{t_{Edd}} = \gamma(\lambda,\epsilon)\frac{dt}{ t_{Edd}}
\end{equation}
where $\gamma(\lambda,\epsilon)$ is the growth efficiency factor. However, because the two parameters in the $\gamma$ function varies with time, we cannot integrate Eqn.~\ref{eqn:diffEqn} unless we assume they are constant (the steady accretion flow scenario) or that they vary very slowly within a time scale of the order of $t_{Edd}$. Assuming the values are constant, integrating Eqn.~\ref{eqn:diffEqn} is straightforward, leading to the solution:
\begin{equation}\label{eqn:growth_curve1}
\frac{M_{BH}}{M_0} = \exp\{\gamma(\lambda,\epsilon)\frac{(t-t_0)}{ t_{Edd}}\}   \end{equation}
where the two constants of integration, $M_0$ and $t_0$, can be interpreted as the mass of the SMBH seed and the cosmological time when the seed starts growing by accretion. 

It was this solution that \citet{Aggarwal2025} adopted to explain the fast growth of SMBHs at different redshifts. This solution also predicts that as the redshift decreases, $\epsilon$ increases while $M_0$ decreases, a result that suggests the accretion phase on SMBH evolved with time, following the cosmological decrease of gas in the intergalactic medium. 

As a consequence of this apparent evolution a drastic change in the formation/evolution of galaxies would be expected between $6<z<7$, suggesting a separate phase in the growth of SMBH in QSOs at these redshifts. However, there is an obvious difficulty with the above model which is that the growth efficiency factor, $\gamma(\lambda_{Edd},\epsilon)$, varies with time at the same rate as the accretion of gas onto the BH decreases:
\begin{equation}\label{gamma1}
\gamma(t) = \lambda_{Edd}(t) \frac{1-\epsilon(t)}{\epsilon(t)} \propto \frac{\dot M(t)}{M(t)}   
\end{equation} 
Consequently, in order to integrate Eqn.~\ref{eqn:diffEqn}, one must first determine what mathematical form has the function $\gamma(t)$, which, after integration, would lead to the solution:
\begin{equation}\label{eqn:growth_curve2}
\frac{M_{BH}}{M_0} = \exp\{\frac{1}{t_{Edd}}\ \int_{t_0}^{t}\gamma(t)dt \}  = \exp\{\frac{\Omega(t)}{t_{Edd}}\}
\end{equation}
where $\Omega(t)$ is the integral of the growth efficiency factor $\gamma(t)$.

\subsection{Optimizing the accretion process}

Although the exact mathematical form of the function $\gamma(t)$ is unknown, we know that the two parameters in Eqn.~\ref{gamma1} are constrained physically to values $\leq 1$. This implies that their variations in time must be coupled. Moreover, we also know that the exponential growth in mass of BHs cannot continue indefinitely but must gradually decrease and eventually stop due to the rapid depletion of gas falling onto the SMBH. This implies that in order to transform $\gamma(t)$ into a genuine growth efficiency function, the accretion of the mass onto the BH must be constrained. The simplest way to do so is to optimize $\dot M$. 

The optimization process begins by deriving Eqn.~\ref{eqn:growth_curve2} once, obtaining $\dot M$, then deriving a second time, applying the optimization condition $\ddot M = 0$. on the other hand, once we derived Eqn.~\ref{eqn:growth_curve2}, we obtain back $\gamma(t)$, which implies that the second derivative is $\dot \gamma(t)$. Therefore, to get $\dot M/M$ we must first determine how the integral of $\gamma(t)$ appearing in Eqn.~\ref{eqn:growth_curve2} is connected to its derivatives. Unfortunately, this information is missing. However, one can guess what is missing by examining how the coupled parameters in $\gamma(\lambda,\epsilon)$ are expected to vary with time. This was done previously by \citet{Aggarwal2025}, who commented that since both values $\lambda$ and $\epsilon$ must decrease with time, $\gamma(\lambda,\epsilon)$ is expected to decrease as $1/t$. This, in fact, is the simplest, most straightforward or direct, solution. 

This suggests that the integral of $\gamma(t)$ could have the form $(t-t_0)/t$. Consequently, we  parameterize the exponent to be optimized in Eqn.~\ref{eqn:growth_curve2} using the function 
$\frac{\Omega(t)}{t_{Edd}}=f(t)= a (t-t_0)/t$, where $a$ is a renormalization parameter that depends on the optimization. The equation to optimize is thus : 
\begin{equation}\label{eqn:growth_curve3}
\frac{M_{BH}}{M_0} = \exp\{\frac{\Omega(t)}{t_{Edd}}\}=\exp\{f(t)\}
\end{equation}

Deriving Eqn.~\ref{eqn:growth_curve3} one gets for the mass accretion $\dot M(t)  = M_0 \dot f(t) e^{f(t)}$, with the optimization condition $\ddot M(\tau)=0$, where $\tau$ is the optimal time, that is, the time at which the accretion reaches its maximum. The optimization condition is thus equal to:
\begin{equation}\label{opt1}
M_0 \ddot f(\tau) e^{f(\tau)} = - M_0 [\dot f(\tau)]^2 e^{f(\tau)} 
\end{equation}
that is:
\begin{equation}\label{opt2}
\ddot f(\tau)  = - [\dot f(\tau)]^2
\end{equation}
Using Eqn.~\ref{opt2} we can now determined the specific value of $a$: 
\begin{equation}\label{apar}
 \frac{a^2 t_0^2}{\tau^4} = \frac{2a t_0}{\tau^3} \Rightarrow a = 2\tau/t_0   
\end{equation}
obtaining for the optimal evolution of the mass growth of the SMBH the expression:
\begin{equation}\label{masstau}
M(t) = M_0 \exp\{(\frac{2\tau}{t_o})(1-\frac{t_0}{t})\}
\end{equation}

To test our solution, we applied the optimal model to a sample of 107 QSOs in the list of \citet{Aggarwal2025} for which the SMBH masses were estimated spectroscopically using the MgII emission line, and a few (16) based on CIV or CII. The SMBH masses varied around $\log M_{BH}= 9\pm 1$ (see Figure~\ref{parametrization_sample}d) and are located within the redshift range $5.5\leq z < 8$. To keep the values of the seed masses above $10^3$~M$_\odot$, in agreement with seed models \citep{Klessen2023,Sanati2025}, we lower $t_0$, that is, the time at which accretion onto the SMBH seeds starts, to a value corresponding in redshift to $z = 24$  (Aggarwal assumed $z=30$). 

\begin{figure*}
\centering
\includegraphics[width=1 \linewidth]{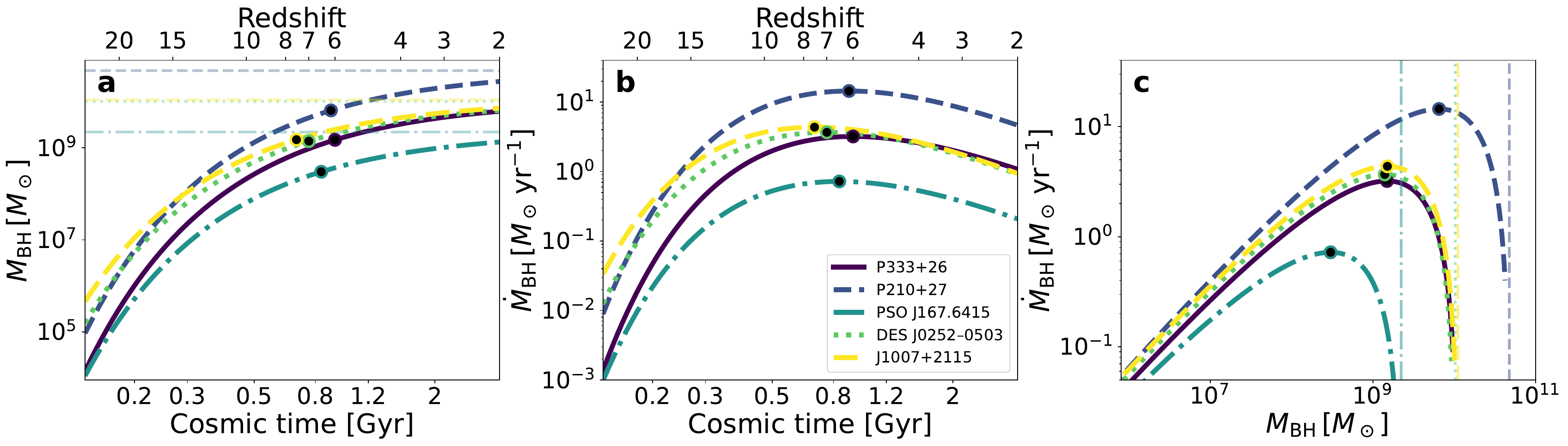}
\caption{Optimized growth functions for five SMBHs at high $z$ listed by \citet{Aggarwal2025}. Panels (a) and (b) show, respectively, their masses at the time of observation, $M_{BH}$, and accretion rates, $\dot{M}_{BH}$, as functions of the cosmic time. Panel (c) shows the growth efficiency curve describing the optimized evolutionary path for each SMBH. The vertical lines in (c) locate the critical mass, $M_c$, which is the mass when the accretion rates asymptotically reach zero.}
\label{optimized_functions}
\end{figure*}

\begin{figure}
\centering
\includegraphics[width=0.9 \linewidth]{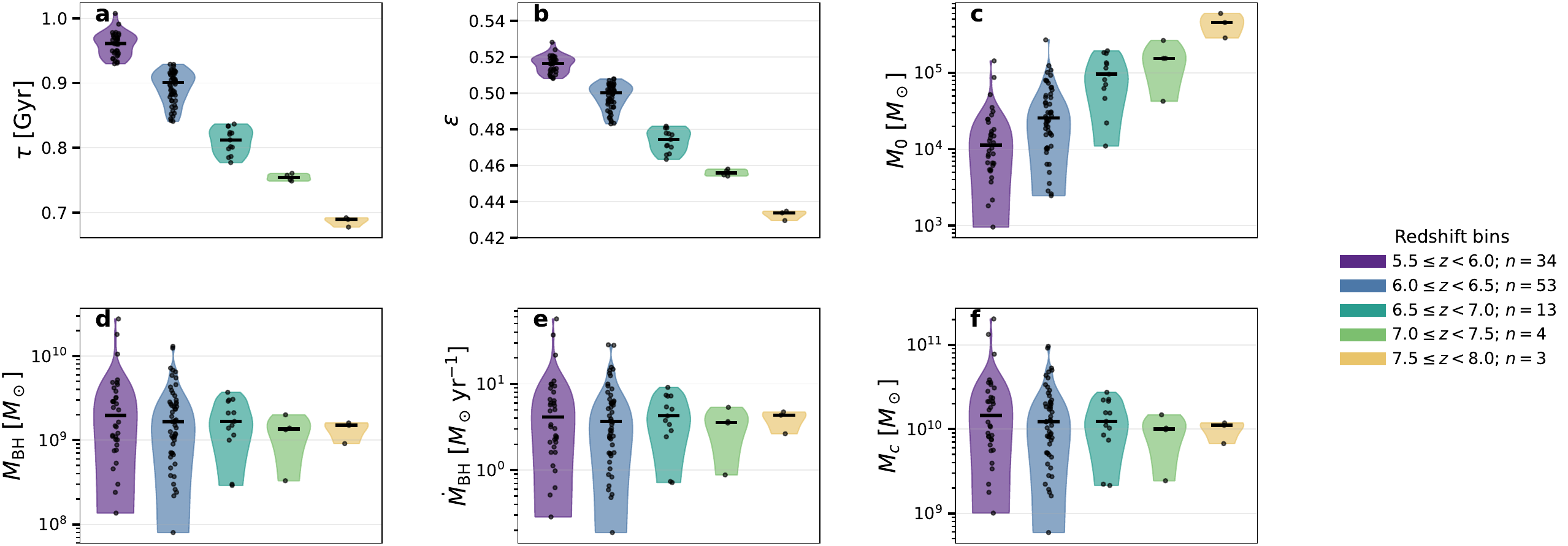}
\caption{Distribution of optimization parameters of the black hole growth for the whole sample of high-redshift galaxies separated in five redshift bins $\Delta z=0.5$ from $z=5.5$ (left) to $z=8.0$  (right).  Optimal time, $\tau$, Radiative efficiency, $\epsilon$, mass of the seed $M_0$, optimal mass, $M_\mathrm{BH}$, maximal accretion rate, $\dot{M}_\mathrm{max}$ and critical mass $M_c$, The dark bars on each plot show the median of the distributions.}
\label{parametrization_sample}
\end{figure}

The solutions of the optimal accretion model are shown in Figure~\ref{optimized_functions}a for five average SMBHs within five redshift bins as identified in the legend of Figure~\ref{parametrization_sample}. The optimal model easily reproduces the high masses of all AGN, without the need for supermassive seeds ($M_0 < 10^6$ M$_odot$) or super-Eddington accretion, $\lambda_{Edd}$. However, experimenting with different values for $\lambda_{Edd}$, we confirm the need for high accretion rates as suggested by different authors, with values within the range $0.6\leq\lambda_{Edd} \leq 1$.

The accretion rate varies as:
\begin{equation}\label{dotmasstau}
\dot M(t) = M_0 \frac{2\tau}{t^2}\exp\{(\frac{2\tau}{t_o})(1-\frac{t_0}{t})\}
\end{equation}
Its variation as a function of cosmic time is traced in Figure~\ref{optimized_functions}b. The optimal value happens at the time we observe the SMBH because this time also marks the moment the accretion begins to fall rapidly to zero in Figure~\ref{optimized_functions}c, which traces the optimized mass growth efficiency curves (MGECs) at different redshifts. 

Each BH has its own MGEC, with proper optimal time and BH seed at redshift $z=24$. This implies that although the growth process is the same for each SMBH the details how they form must depend not only on the formation of the SMBH seeds but also on the primordial conditions of accretion, that is, primordial local gas densities and environments where the host galaxies formed. 

According to the optimal accretion model, the function $\gamma$ has now the form:
\begin{equation}
\gamma(t) =  \frac{2\tau t_{Edd}}{t^2}   
\end{equation}

Which allows us to estimate $\epsilon$ (something no other model can do) using the following relation:
\begin{equation}
\epsilon =  \{\frac{2\tau t_{Edd}}{t^2}\frac{1}{\lambda_{Edd}}+1\}^{-1}  \end{equation}

Assuming $\lambda_{Edd} = 1$ (the only free parameter), the optimal model predicts high values of $\epsilon$, consistent with the high luminosity of QSOs. In Figure~\ref{parametrization_sample}a and b the variation of the optimal time, $t=\tau$, follows the variation in radiative efficiency, the two values decreasing as $z$ increases. This is in good agreement with the idea that the reason why SMBHs grow in mass so fast at high $z$ is because their accretion disks tend to have lower radiative efficiency than at low $z$ \citep{King2006,Aggarwal2025}. 

Finally in Figure~\ref{parametrization_sample}c the optimal model predicts an increase in $M_0$ at high $z$. Taken at face value, this trend could be seen as evidence of downsizing, that is, more massive BH at high $z$ forming before less massive BH at low $z$. However, very few differences are observed between the three optimized parameters involving the mass, $M_{BH}$, $\dot M_{BH}$ and $M_c$, respectively in Figure~\ref{parametrization_sample}d, e and f. Note that $M_{BH}$ is independent of the model, suggesting that there are no significant differences in mass at different redshifts is physical (only an increase in dispersion is observed, consistent with observational biases). Now, since the optimal accretion rate happens by definition at the exact same moment we observe the QSOs, and because the final mass, $M_c$, when the accretion reach zero is reached a short time after this moment, then naturally all the variations related to the mass are proportional to one another. Therefore, it seems that to explain the similarity in masses in the optimal model, the two parameters $\epsilon$ and $M_0$, must vary exactly in the opposite sense relative to the redshift. This suggests that despite the conditions for their formation being varied, the end product of the formation processes of SMBH at any redshift is essentially the same. In other words, the solution is unique, a statement which looks strangely reminiscent of the no hair theorem \citep{Israel1967,Bekenstein1980}.



\subsection{Connection with standard thermodynamics}

\begin{figure}
\centering
\includegraphics[width=0.9 \linewidth]{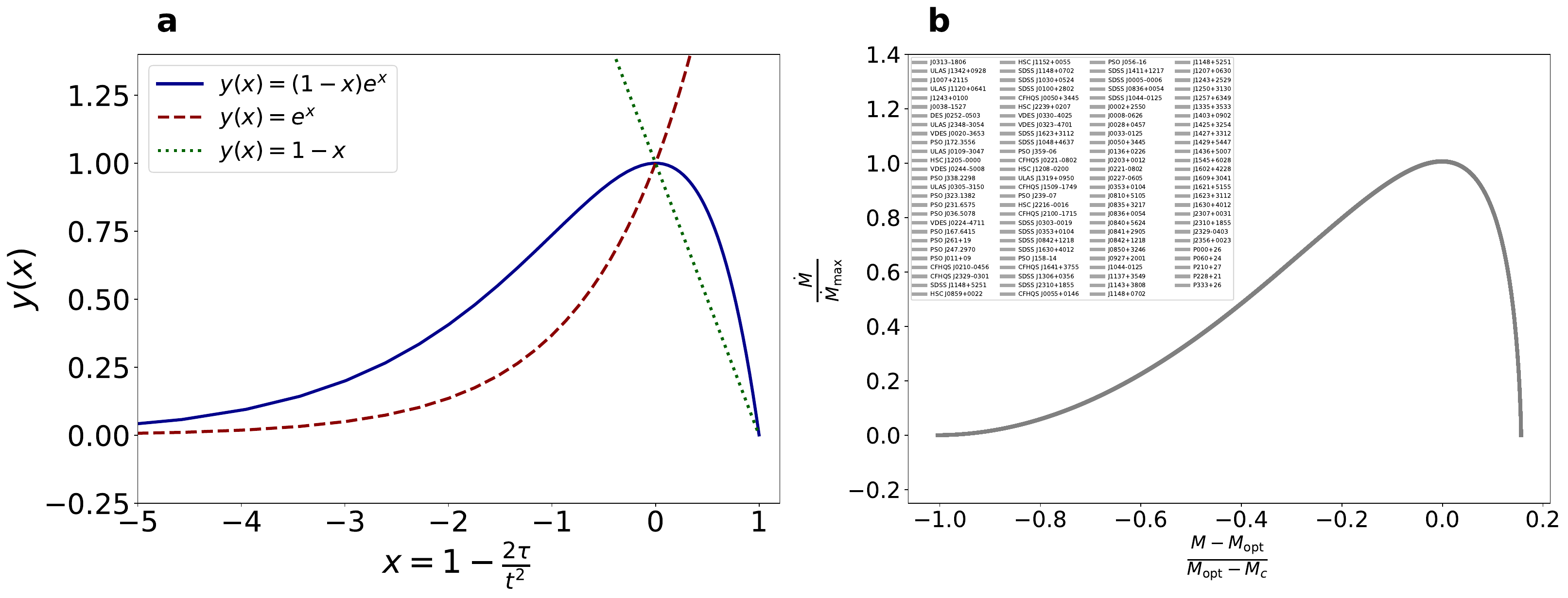}
\caption{Mass growth efficiency curve (MGEC) for SMBHs. Panel a: The time scale was calibrated based on the observations, where $\tau$ is the optimal time. Panel b: This figure is similar as Fig.4 in \citet{Arnoldi2025} replacing the temperatures by the masses. The mass scale is relative to the mass at the optimal time. The 107 SMBHs (identified in the label) trace exactly the same curve.}
\label{performance_curve2}
\end{figure}

A few more explanations about the optimal formation processes of SMBHs are necessaries in order to understand why they lead apparently to a solution that is unique. The accretion of matter onto a BH is optimized, which implies that its exponential growth in mass is limited by the decrease in accretion rates with time. Mathematically the MGEC, in Figure~\ref{optimized_functions}c all have the same generic form (Figure~\ref{performance_curve2}a): an exponential function combined with a linear function that decreases with time, where the calibration of the time scale corresponds to the term multiplying the exponent in Eqn.~\ref{dotmasstau}. This term, $2\tau/t^2$, corresponds to the growth efficiency per mass, $\dot M/M$ (dividing Eqn.~\ref{dotmasstau} by Eqn.~\ref{masstau}), which, when $t=\tau$, is consistent with the decreasing part of the generic function $(1-x)$. This is the function that limits the growth of the BH (the green dot line).

Consequently, the SMBHs observed correspond in Figure~\ref{optimized_functions}a to the mass of the SMBHs at the optimal time in Figure~\ref{optimized_functions}b, because this phase in the growth history process corresponds to the state with the highest probability of detection: 
this happens when the accretion rate is maximum, that is, when the SMBH is most active, and thus easier to observe, at the peak of the MGEC in Figure~\ref{optimized_functions}c. 

As it tuned out, the generic MGEC for SMBHs (Figure~\ref{optimized_functions}a) has the exact same mathematical form as the Temperature Performance Curve (TPC) recently discovered by \citet{Arnoldi2025} in biology, which describes ``how rates of performance or fitness traits change as a function of the temperature across a wide range of living systems.'' According to \citet{Arnoldi2025} the TPC is universal, which implies that despite the variety and complexity of the mechanisms explaining the behaviours of biological systems, their performance curves always converge to the UTPC. In other words, the temperature in the UTPC acts as a thermodynamics state function, whose value depends only on the current equilibrium state of the system, not on the path or processes taken to reach that state.  

This is also true in the case of SMBHs. In Figure~\ref{performance_curve2}b, which is similar to Fig.~4 in \citet{Arnoldi2025} but where the mass replaces the temperature, all the SMBHs in our sample trace exactly the same curve; there are exactly 107 curves traced in Figure~\ref{performance_curve2}b that fall precisely on each other. In other words, the mass in the MGEC is equivalent to a thermodynamics state function, which explains why despite their different histories of formation and BH seeds the final states of the growth in mass of SMBHs are all the same. 


In \citet{Bekenstein1972} and \citet{Bardeen1973} similar analogies between BH mechanics and thermodynamics were already noted and discussed in terms of generalized thermodynamics. However, the analogy between the UTPC and MGEC seems to go deeper, suggesting that the mass growth process of SMBHs is equivalent to a standard thermodynamics process, where the mass is a state function.

\end{document}